\documentclass[aip,jcp,reprint,floatfix]{revtex4-1}  
\pdfoutput=1

\usepackage[T1]{fontenc}

\usepackage{fixmath}
\usepackage{times}
\usepackage[slantedGreek]{mathptmx}
\usepackage[Symbol]{upgreek}
\usepackage[dvipsnames]{xcolor}
\usepackage{geometry}
\DeclareSymbolFont{UPM}{U}{eur}{m}{n}
\DeclareMathSymbol{\partial}{0}{UPM}{"40}

\usepackage{helvet}
\usepackage{graphicx}
\usepackage{amsmath,amssymb}

\usepackage[stretch=15,shrink=15,step=3]{microtype}
\usepackage{cmap}
\usepackage{siunitx}
\usepackage[nodayofweek]{datetime}
\newdateformat{myDate}{\THEDAY\ \monthname[\THEMONTH] \THEYEAR}

\usepackage{hyperref}
\hypersetup{colorlinks, citecolor={blue},linkcolor={red}, urlcolor={violet},
pdftitle={Homogeneous TIP4P/2005 ice nucleation at low supercooling},
pdfauthor={Aleks Reinhardt, Jonathan P. K. Doye}, pdfdisplaydoctitle}

\begin{document}

\title{Homogeneous TIP4P/2005 ice nucleation at low supercooling}
\author{Aleks Reinhardt}
\author{Jonathan P.~K.~Doye}
\affiliation{Physical and Theoretical Chemistry Laboratory, Department of
Chemistry, University of Oxford, Oxford, OX1 3QZ, United Kingdom}
\date{21 June 2013}




\maketitle

Although homogeneous ice nucleation is thought to be an important process in
atmospheric science\cite{Oxtoby1992, *Baker1997, *Sassen2000, *Hegg2009,
*Murray2012, *Khvorostyanov2012} and understanding how ice grows has recently
been identified as one of the top open questions in ice
science,\cite{BartelsRausch2013} simulating the process has been fraught with
difficulties.\cite{Radhakrishnan2003b, *Radhakrishnan2003, *Quigley2008,
Brukhno2008, *Moore2011, *Li2011, *Cox2013, Reinhardt2012, Reinhardt2012b} This
is especially the case for all-atom model simulations, since water dynamics at
significant supercoolings are very slow indeed,\cite{Reinhardt2012b} which has
made the determination of free energy landscapes and nucleation rates using such
models very difficult.\cite{Reinhardt2012b} One approach that could provide some
insight into the process, while still being computationally tractable, would be
to simulate homogeneous nucleation using the TIP4P/2005 model of
water\cite{Abascal2005} at temperatures at which the dynamics are reasonably
fast and at which equilibration can thus be achieved, at least for relatively
small crystalline nuclei.

\begin{figure}[tbp]
\begin{center}
\includegraphics{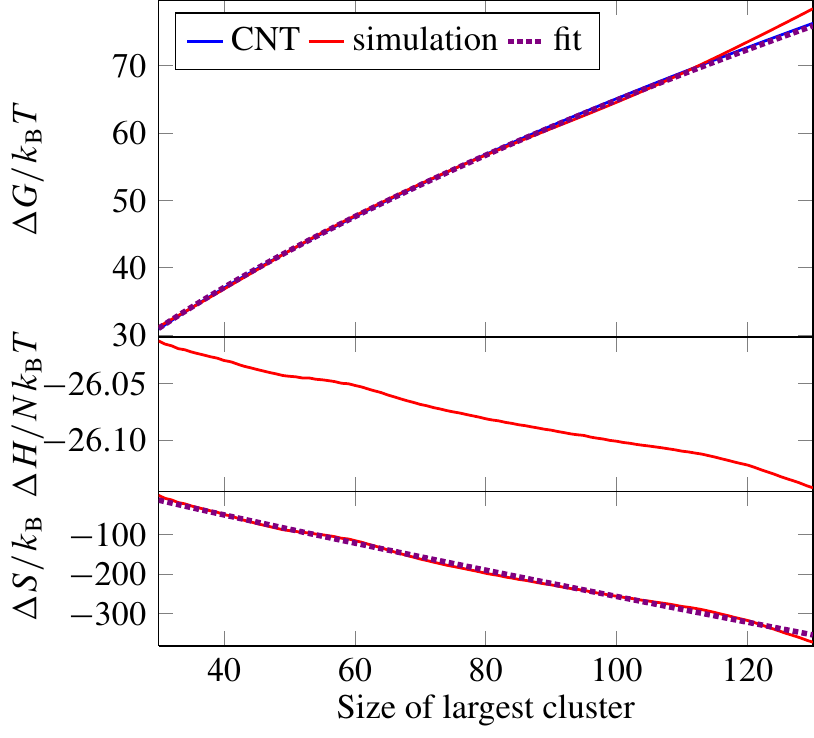}
\end{center}
\caption{The Gibbs energy, enthalpy and entropy for TIP4P/2005 ice nucleation as
a function of the size of the largest crystalline cluster for systems seeded
with a cubic ice seed. The hexagonal nucleation seed simulation results are
analogous. $T=\SI{240}{\kelvin}$, $p=\SI{1}{\bar}$, $N=2500$ molecules. In the
top panel, the simulation result is shifted so that it matches the CNT estimate
at size 50, and this estimate is also shown. In the bottom panel, the curve is
shifted so the simulation $\upDelta S$ is zero at a size of
30.}\label{fig-TIP4P-HMC-weights-data-240K}
\end{figure}

We have run simulations analogous to those presented in
Ref.~\citenum{Reinhardt2012b} using hybrid Monte Carlo\cite{Duane1987} with
adaptive umbrella sampling\cite{Torrie1977, *Mezei1987} and a local order
parameter to drive the process.\cite{Reinhardt2012b} These simulations started
with seed hexagonal and cubic ice crystals at \SI{240}{\kelvin} and
\SI{1}{\bar}, which is a $\sim$\SI{5}{\percent} supercooling for TIP4P/2005
water. The starting umbrella weights corresponded to the negatives of the free
energy barrier estimated from classical nucleation theory
(CNT).\cite{Anwar2011b} A free energy profile for nucleation from such
simulations is shown in Fig.~\ref{fig-TIP4P-HMC-weights-data-240K}. Whilst the
range of crystalline cluster size presented here is rather limited
(\textit{e.g.}, the critical cluster predicted by CNT is of the order of
\num{1.5e4} molecules), even the calculation of just this set of free energies
represents a huge computational effort, as even at such small supercoolings, the
dynamics of ice growth are slow on computational time scales.  Consequently,
obtaining more complete data would be prohibitively expensive, particularly for
larger cluster sizes for which larger system sizes would need to be simulated.

To test whether equilibrium has been attained, we need to ensure that there is
sufficient sampling and frequent exchange across order parameter values. We can
also calculate the enthalpy as a function of the order parameter and ensure that
it behaves sensibly. These equilibration criteria are fulfilled by the
simulations whose results are presented in
Fig.~\ref{fig-TIP4P-HMC-weights-data-240K}, and so we have obtained equilibrium
results for ice nucleation at low supercoolings. We remark that the reason for
the deviation from the trend at the extreme cluster size values is due to
windowing errors, which are especially significant when instantaneous order
parameter values can fluctuate to the extent that they can in these
systems.\cite{Reinhardt2012b}

If we fit the simulation data shown in
Fig.~\ref{fig-TIP4P-HMC-weights-data-240K} to a CNT-like expression for the
Gibbs energy, as we did when studying ice nucleation\cite{Reinhardt2012} using
the mW monatomic model of water,\cite{Molinero2009} we find that
$\upDelta_\text{fus}\mu(\SI{240}{\kelvin})/k_\text{B}\approx\SI{29.3}{\kelvin}$
and $\gamma\approx \SI{24.0}{\milli\joule\per\metre\squared}$; these values
compare favourably to the
$\upDelta_\text{fus}\mu(\SI{240}{\kelvin})/k_\text{B}\approx\SI{27.7}{\kelvin}$
estimated using the approximation that $N\upDelta_\text{fus}\mu \approx
\upDelta_{\text{fus}} H \left(  1 - T/T_\text{fus} \right)$ and $\gamma\approx
\SI{24.5}{\milli\joule\per\metre\squared}$ calculated for the basal plane of
TIP4P ice.\cite{Davidchack2012} The simulation results appear to agree very well
with the CNT prediction; indeed, the agreement is almost suspiciously good.
Given that we started the simulations with umbrella weights corresponding to
CNT, we could envisage a situation where the clusters essentially remain at
their original size because of the slow dynamics of the ice cluster
growth/shrinkage process. While this is certainly always a conceivable issue in
simulations of water, it does not appear to be a problem in these
high-temperature simulations, since in equivalent simulations with starting
weights corresponding to CNT weights at \SI{235}{\kelvin} and at
\SI{245}{\kelvin} (with the simulation temperature remaining at
\SI{240}{\kelvin}), clusters were observed to shrink over time with the former
and to grow with the latter set of umbrella weights. This suggests that the CNT
estimate at the temperature of simulation does in fact yield a reasonable
approximation to the free energy barrier associated with the simulated process.
However, the good agreement in this very early stage of the nucleation process
is no guarantee that the critical cluster size or the height of the free energy
barrier to nucleation are also well estimated by CNT.

It is also interesting to note that the enthalpy associated with the growth of
the crystal nucleus is a monotonically downhill function of the nucleus size
(Fig.~\ref{fig-TIP4P-HMC-weights-data-240K}); this result implies that,
analogously to the mW simulation results,\cite{Reinhardt2012} the barrier to
nucleation is primarily entropic in nature. This suggests that the formation of
an ice-liquid surface is enthalpically favourable (or at least neutral), but
entropically unfavourable. Since $(\partial \gamma / \partial T)_p =
-S_\text{interface}$, where $S_\text{interface}$ is the entropy change per unit
area upon the formation of an interface, this finding is consistent with
experiment, where the interfacial free energy was found to decrease with
temperature.\cite{Wood1970, *Huang1995, *Murray2010} We can attempt to quantify
this temperature dependence to first order by explicitly evaluating this
interfacial entropy. To do this, we first find $\upDelta S(n)$, where $n$ is the
number of molecules in the largest crystalline cluster, by using $\upDelta G(n)$
and $\upDelta H(n)$ determined from the simulation
(Fig.~\ref{fig-TIP4P-HMC-weights-data-240K}). We then proceed to fit these
values to $\upDelta S(n) = a n + b n^{2/3} + c n^{1/3} + d$, where $a$ is
constrained to be the bulk entropy change at coexistence (namely
$a=\upDelta_\text{fus} S/N = \upDelta_\text{fus} H / NT_\text{fus}$, where
$T_\text{fus}=\SI{252}{\kelvin}$), and we assume that $b =S_\text{interface}
(36\uppi/\rho_\text{ice}^2)^{1/3}$, where the factor
$(36\uppi/\rho_\text{ice}^2)^{1/3}$ accounts for the assumed spherical shape of
the clusters. This gives a value of $S_\text{interface}$ of
\SI{-0.18}{\milli\joule\per\square\metre\per\kelvin}, and also allows us to
estimate the interfacial enthalpy, defined by
$H_\text{interface}=\gamma+TS_\text{interface}$, as
\SI{-18.6}{\milli\joule\per\square\metre}. Integrating $(\partial \gamma /
\partial T)_p = -S_\text{interface}$ with respect to the temperature, assuming
that the entropy is independent of temperature, gives an interfacial free energy
at coexistence of \SI{26.1}{\milli\joule\per\square\metre}. This is reasonably
consistent both with the mean values obtained by Davidchack and co-workers for
TIP4P water,\cite{Davidchack2012} as well as the TIP4P/2005 estimate of
$\gamma\approx \SI{28}{\milli\joule\per\metre\squared}$ obtained from CNT
critical cluster size fits for small supercoolings.\cite{Sanz2013} The relative
agreement between these different approaches to obtaining $\gamma$ is
interesting, given the contrasting behaviour for systems such as
NaCl.\cite{ZykovaTiman2008} However, it should be borne in mind that our
estimates are rather crude, as we (a) assume that classical nucleation theory
applies, (b) calculate non-linear fits to the data, where a variety of fits is
likely to lead to reasonable agreement, and (c) extract these data from a
relatively small range of cluster sizes.

In their work, Limmer and Chandler calculate an estimate of the interfacial free
energy and its variation with temperature for the mW model of
water.\cite{Limmer2012} They demonstrate that for the mW model, the Turnbull
relation, $\gamma(T_1)/ \upDelta_\text{fus} H(T_1) = \gamma(T_2)/
\upDelta_\text{fus} H(T_2)$,\cite{Turnbull1950} works remarkably well. If we
assume that the same relation applies to TIP4P/2005 water and that
$\gamma(\SI{240}{\kelvin})$ as reported above is correct, and we obtain
$\upDelta_\text{fus} H(T)$ from fits to the internal energy and density of ice
and liquid water found in the literature,\cite{Noya2007b, *Pi2009} we find that
$\gamma(\SI{252}{\kelvin})\approx\SI{27.5}{\milli\joule\per\metre\squared}$,
again in reasonable agreement with our estimate for this temperature.

In summary, we have presented the first results for the free energy profile
associated with the homogeneous nucleation of ice using a rotationally invariant
local orientational order parameter for an all-atom model of water. These low
supercooling results corroborate our hypothesis\cite{Reinhardt2012,
Reinhardt2012b} that for ice nucleation, classical nucleation theory predictions
may be considerably better than might initially be assumed; this is consistent
with the results of Ref.~\citenum{Sanz2013}. We note that the free energy
barriers obtained in some previous simulations,\cite{Radhakrishnan2003b,
*Radhakrishnan2003, *Quigley2008} which were significantly larger than the CNT
estimate, likely arise from the use of global order parameters, which can result
in nucleation pathways that are not the lowest in free energy in systems with
slow dynamics.\cite{Reinhardt2012b} Furthermore, we have shown here that the
interfacial entropy is negative: presumably, the hydrogen bonding of liquid
molecules with the ice nucleus considerably constrains the hydrogen bond network
in the liquid near the surface. However, our attempts to determine the overall
free energy barrier and nucleation rate of homogeneous ice nucleation using
all-atom models of water have been thwarted by the slow dynamics of the system.
We note that despite our attempts to circumvent this difficulty, such as using
histogram reweighting and hamiltonian exchange,\cite{Reinhardt2013b} we have
been unable to obtain simulation results at sufficiently low temperatures to
enable us to calculate nucleation rates. The slow dynamics of ice growth at low
temperatures thus continue to pose a very significant obstacle. The use of
advanced simulation methods does offer new insights into the process;
nevertheless, the successful calculation of ice nucleation rates for all-atom
models of water continues to be a challenge.

We thank the EPSRC for financial support.

%

\end{document}